# Quantitative evaluation of outdoor artificial light emissions using low Earth orbit radiometers

Salvador Bará[1,*], Carmen Bao-Varela[2], and Raul C. Lima[3,4]

[1] A. Astronómica 'Ío', 15005 A Coruña, Galicia

[2] Photonics4Life Research Group, Applied Physics Department, Universidade de Santiago de Compostela, Campus Vida, E-15782 Santiago de Compostela, Spain

[3] Escola Superior de Saúde do Instituto Politécnico do Porto, 4200-072 Porto, Portugal

[4] Instituto de Astrofísica e Ciências do Espaço (IA), Univ Coimbra, 3040-004, Coimbra, Portugal

(*) email: salva.bara@usc.gal

**Abstract**

Low Earth orbit radiometers allow monitoring nighttime anthropogenic light emissions in wide areas of the planet. In this work we describe a simple model for assessing significant outdoor lighting changes at the municipality level using on-orbit measurements complemented with ground-truth information. We apply it to evaluate the transformation effected in the municipality of Ribeira (42°33'23"N, 8°59'32" W) in Galicia, which in 2015 reduced the amount of installed lumen in its publicly-owned outdoor lighting system from 93.2 to 28.7 Mlm. This significant cutback, with the help of additional controls, allowed to reduce from 0.768 to 0.208 Mlm/km$^2$ the lumen emission density averaged across the territory. In combination with the VIIRS-DNB annual composite readings these data allow to estimate that the relative weight of the emissions of the public streetlight system with respect to the total emissions of light in the municipality changed from an initial value of 74.86% to 44.68% after the transformation. The effects of the sources' spectral shift and the photon calibration factor on the radiance reported by the VIIRS-DNB are also evaluated.

# 1. Introduction

The sustained increase of the nighttime emissions of artificial light is a subject of growing concern for its unwanted effects on biodiversity, sustainability of scientific research, preservation of sky-related intangible heritage, nocturnal nightscapes, and, potentially, public health [1-3]. A combination of instrumental techniques are nowadays available for monitoring this process from local to global scale, including permanent networks of ground-based nighttime radiometers [4-7], observation campaigns with different types of detectors [8], and airborne [9-12] and low Earth orbit (LEO) satellite platforms [13-21].

LEO radiometers are suitable instruments for monitoring nighttime lights in large territories with spatial resolution ranging from hundreds of meters to meters, temporal resolution of order one day and spectral capabilities from panchromatic to moderately multispectral. Several products from different orbital platforms are publicly available for general purpose applications [16-17,20]. Some of them have been used for estimating the relative contribution of the public outdoor lighting systems to the total light emissions at the city or municipality levels [22-23], complementing ground-based methods based on the analysis of the light scattered by the atmosphere [24-25]. These studies provided divergent estimates of the contribution of the public lighting systems to the total nighttime light emissions, a not unexpected situation that can be traced back to the specific schemes of ownership and use of light sources in different socio-political realities.

It is the purpose of this paper to assess the feasibility of quantitatively monitoring municipality-wide, large outdoor lighting changes using satellite time-averaged composites of nighttime lights, complemented by reliable ground-truth information. To that end, we first describe a basic radiometric model relating the changes in overall lumen emissions to the changes in the radiance reported by an on-orbit radiometer, accounting for the specific spectral shifts arising from the substitution of gas-discharge lamps by solid-state light emitting diodes (LED). For the subsequent practical application the model will explicitly refer to the VIIRS-DNB aboard the Suomi-NPP satellite [14, 20], but with the appropriate modifications in the detector passband and photon-to-radiance conversion factor is applicable to any other orbital remote sensing platform working in the optical region of the spectrum. This model is applied here to analyze the evolution of the Galician municipality of Ribeira, which in 2015 carried out a substantial remodeling of its public lighting system, with a drastic reduction in the nominal installed lumen, from 93.2 to 28.7 Mlm. This information, in combination with the satellite measurements, is used here to estimate the relative weight of the public lighting system of this municipality in the overall nighttime emissions before and after this remodeling operation.

The basic calculation model is outlined in section 2. Section 3 describes the parameters of the public lighting remodeling in Ribeira, the expected and measured changes in the signal recorded by the VIIRS-DNB radiometer in the average yearly composites [20], and the information that can be extracted from these data. Discussion and conclusions are summarized in sections 4 and 5, respectively.

## 2. Methods: Top-of-atmosphere radiance from urban light sources

Most present-day streetlights belong to a small set of basic lamp technologies [26-28], composed of black-body radiators, gas-discharge and solid-state lighting emitters. Their spectral distributions are often available as spectral densities $\Phi(\lambda)$, corresponding to irradiances measured at some reference plane or to radiant fluxes given in arbitrary linear units per nm. These arbitrary-units distributions can be converted to normalized spectral radiant power densities in W·nm$^{-1}$ per emitted lumen (lm), $\widehat{\Phi}(\lambda)$, as

$$\widehat{\Phi}(\lambda) = \frac{\Phi(\lambda)}{683 \int_{\lambda=0}^{\infty} \Phi(\lambda)\, V(\lambda)\, d\lambda} \qquad (1)$$

Where $V(\lambda)$ is the photopic spectral sensitivity function of the human visual system [29], $d\lambda$ is expressed in nm, and 683 lm·W$^{-1}$ is the luminous efficacy of monochromatic electromagnetic radiation of frequency $540 \times 10^{12}$ Hz, a scaling constant central to the SI definition of the basic luminous intensity unit [30], the candela (cd).

*2.1 Ground emissions*

Let us consider a public lighting system composed of $i = 1, \ldots, I$ different types of lamps, each characterized by a specific spectrum $\widehat{\Phi}_i(\lambda)$ normalized to 1 lm, and let the maximum amount of lumen emitted by each individual lamp at the beginning of their lives be $N_{Li}$. During their useful life the average lamp emission will progressively decrease due to several aging mechanisms, a phenomenon known as "lamp lumen depreciation", such that the actual number of emitted lumens at any given point on time will be a fraction $f_i$ of the initial maximum value. On the other hand, especially with black-body and gas-discharge lamps, a non-negligible fraction of the luminous flux will be absorbed by the luminaire enclosure itself, and the amount of lumen actually leaving the luminaires will be a fraction $g_i$ of the lumen emitted by the lamps, fraction known as "luminaire efficiency". Furthermore, the streetlight emissions are often purposedly reduced, in some cases by a substantial amount, during those periods of the night when the light demand decreases, by applying an energy-saving regulation factor $r_i$. Assuming there are installed $n_i$ individual lamps of each type, the spectral radiant flux $\Phi_T(\lambda)$ in W·nm$^{-1}$ emitted by the whole lighting system can be expressed as:

$$\Phi_T(\lambda) = \sum_{i=1}^{I} n_i \, f_i \, g_i \, r_i \, N_{Li} \, \widehat{\Phi}_i(\lambda) \qquad (2)$$

Let us denote by $u_i$ the fraction of this radiation that is emitted directly to the atmosphere without interacting with neighboring urban surfaces. This fraction is often of the same order of magnitude, although not exactly equal, to the upward light output ratio (*ULOR*), the fraction of light emitted by individual luminaires in directions above the horizontal. For LED-based luminaires without diffusing screens the *ULOR* tends to be very small, even close to zero due to the well-defined directionality of this type of light sources. Luminaires containing black-body or gas-discharge lamps have a larger active light emitting area, being more difficult to collimate, and lamps with diffusers or structural parts reflecting light upwards, may easily have *ULOR* of order 10%-15%. The associated value of $u_i$ will generally be smaller than the *ULOR*, since part of the light emitted above the horizontal is intercepted by urban vertical surfaces (façades and walls) where it undergoes diffuse reflections, falling outside the $u_i$ fraction definition. According to these definitions, the spectral flux from the luminaires that will interact with the surrounding surfaces, $\Phi_s(\lambda)$, is

$$\Phi_s(\lambda) = \sum_{i=1}^{I} n_i\, f_i\, g_i\, r_i\, (1-u_i) N_{Li}\, \widehat{\Phi}_i(\lambda) = F_S \sum_{i=1}^{I} \gamma_i\, \widehat{\Phi}_i(\lambda) \qquad (3)$$

where $F_S$ is the total amount of lm emitted by the luminaires towards the surfaces and $\gamma_i$ is the fraction of $F_S$ corresponding to each lamp type.

Denoting by $S$ the total area of the surfaces on which this flux is incident, the average spectral irradiance on these surfaces is $E_s^-(\lambda) = \Phi_s(\lambda)/S$ (W·m$^{-2}$·nm$^{-1}$). The angular distribution of the reflected radiance depends on the bidirectional reflectance distribution function (*BRDF*) of each surface patch. Most public streetlights are located directly above artificial surfaces composed of different types of concrete, asphalt, bricks, and stone, in some cases coated by paints and surrounded by vertical surfaces made of the same type of materials. A relatively small fraction of all urban streetlights is directly located over vegetal soils. As a simplifying assumption, the average *BRDF* can be considered approximately Lambertian, with a spatially homogeneous but wavelength-dependent spectral reflectance $\rho(\lambda)$. In such case $L_s^+(\lambda)$, the average ground-level reflected spectral radiance propagating upwards in the direction of the orbital radiometer, usually given in W·m$^{-2}$·sr$^{-1}$·nm$^{-1}$, can be expressed as

$$L_s^+(\lambda) = \frac{\rho(\lambda)}{\pi} E_s^-(\lambda) = \frac{\rho(\lambda)}{\pi\, S} F_S \sum_{i=1}^{I} \gamma_i\, \widehat{\Phi}_i(\lambda) \qquad (4)$$

In most present-day public streetlight systems the fraction $u_i$ of light not interacting with surfaces propagates in angles close to the horizontal, not directly upwards. Only a small fraction of this light gets redirected toward near-zenith angles after being scattered by the atmospheric molecular and aerosol constituents. This scattered radiance is detectable by on-orbit radiometers [31] and contains useful information about the aerosol properties [32-33], but its absolute value in highly lit places [34] tends to be significantly smaller than that of the direct radiance and will not be included in this simplified model.

## 2.3. Top-of-atmosphere radiance

The average ground-level radiance in Eq.(4) will be attenuated before reaching the top of the atmosphere (*TOA*), due to atmospheric extinction (by absorption, and scattering out of the beam) and potential blocking by obstacles, e.g. vegetal canopies, or shadow areas in streets of large shape factor for nonzero nadir satellite observation angles, see [34-36]. Since these two attenuation processes are generally independent from each other, the overall transmittance factor ground to *TOA* can be factored in a term $T_{atm}(\lambda, \mathbf{a})$ accounting for the atmospheric attenuation, and a term $Q(\mathbf{a})$ accounting for shadowing due to obstacles, where $\mathbf{a} = (z, a)$, being $z$ the zenith emission angle (approximately equal to the satellite nadir observation angle, for not very large scanning swaths), and $a$ the azimuth of the satellite as seen from the light source. For a layered atmosphere the attenuation term is independent from the azimuth, and its associate transmittance is $T_{atm}(\lambda, z) = \exp\{-M(z)[\tau_M(\lambda) + \tau_A(\lambda)]\}$, where $M(z)$ is the number of airmasses, that can be approximated to within a 1% by $M(z) = 1/\cos(z)$ for zenith angles up to $z \approx 60°$ (more precise values up to the horizon can be computed with suitably extended expressions see e.g. eq.(30) in Masana et al [37], after Kasten and Young [38]), and $\tau_M(\lambda)$ and $\tau_A(\lambda)$ are the molecular (Rayleigh, ROD) and aerosol (AOD) optical depths, respectively. These optical depths are wavelength-dependent and can be calculated as

$$\tau_M(\lambda) = 0.00879 \, \lambda^{-4.09} \qquad (5)$$

with $\lambda$ in micron [39], and

$$\tau_A(\lambda) = \tau_{A,0}(\lambda_0) \left(\frac{\lambda}{\lambda_0}\right)^{-\alpha} \qquad (6)$$

where $\tau_{A,0}(\lambda_0)$ is the AOD for a given $\lambda_0$ taken as reference and $\alpha$ is the Angström exponent ($\alpha \approx 1$). The obstacle transmittance term $Q(\mathbf{a})$, in turn, depends on the specific urban and topographic features of each location; its spatial average over large regions with dispersed small population nuclei can be assumed to be azimuthally symmetric, $Q(\mathbf{a}) \equiv Q(z)$, while for large urban areas with strong directional properties the azimuthal dependence cannot be ignored [35].

The spectral radiance arriving to the satellite (W·m$^{-2}$·sr$^{-1}$·nm$^{-1}$) is then

$$L^+_{TOA}(\lambda, z) = Q(z) \, T_{atm}(\lambda, z) \, \frac{\rho(\lambda)}{\pi \, S} \, F_S \sum_{i=1}^{I} \gamma_i \, \widehat{\Phi}_i(\lambda) \qquad (7)$$

*2.3 VIIRS-DNB detected radiance*

The energy radiance in Eq. (7) corresponds to a VIIRS-DNB detected in-band photon radiance $n_{DNB}(z)$,

$$n_{DNB}(z) = \int_{\lambda=0}^{\infty} T_{DNB}(\lambda)\, L_{TOA}^{+}(\lambda, z) \left(\frac{\lambda}{hc}\right) d\lambda \tag{8}$$

In photons·s$^{-1}$·m$^{-2}$·sr$^{-1}$, where $T_{DNB}(\lambda)$ is the VIIRS-DNB radiometer passband (Fig. 1). The VIIRS-DNB products report the radiance in energy units nW·cm$^{-2}$·sr$^{-1}$ [20, 40], rather than in photon numbers. Recall that the conversion between these two types of units depends on the detailed shape of the light spectrum. Whereas for narrowband detection the photon number and the energy are roughly proportional independently from the input spectra, this does not hold for panchromatic bands like the VIIRS-DNB one (500–900 nm) where radiance differences of order ~10% can arise for the same number of detected photons. The VIIRS-DNB radiometer is calibrated on-orbit by translating the calibrations of lower gain bands obtained using a solar diffuser [41]. It can then be expected that the conversion from detected photons·s$^{-1}$·m$^{-2}$·sr$^{-1}$ to VIIRS-DNB reported band-integrated radiances in W·m$^{-2}$·sr$^{-1}$ is based on the solar spectrum. This means that the reported radiances, $L(z)$, are given by

$$L(z) = K_{DNB}\, n_{DNB}(z) \tag{9}$$

where $K_{DNB}$ is the conversion factor

$$K_{DNB} = \frac{\int_{\lambda=0}^{\infty} T_{DNB}(\lambda)\, E_{Sun}(\lambda)\, d\lambda}{\int_{\lambda=0}^{\infty} T_{DNB}(\lambda)\, E_{Sun}(\lambda) \left(\frac{\lambda}{hc}\right) d\lambda} \tag{10}$$

in reported W per photon·s$^{-1}$. Using for $E_{Sun}(\lambda)$ the STIS_002 extra-atmospheric solar irradiance spectrum in the CALSPEC database [42] this parameter has the value $K_{DNB} = 2.9619 \times 10^{-19}$ W/(photon·s$^{-1}$).

The modelled in-band radiance reported by the VIIRS-DNB can be finally written as:

$$L(z) = \frac{F_S}{S} K_{DNB}\, Q(z) \int_{\lambda=0}^{\infty} T_{DNB}(\lambda)\, T_{atm}(\lambda, z)\, \frac{\rho(\lambda)}{\pi} \sum_{i=1}^{I} \gamma_i\, \widehat{\Phi}_i(\lambda) \left(\frac{\lambda}{hc}\right) d\lambda \tag{11}$$

This radiance can be read as $L(\boldsymbol{\beta}) = N\, \mathcal{H}(\boldsymbol{\beta})$, i.e. as the product of the average spatial density of light emissions $N = F_S/S$ (in lm·m$^{-2}$ or Mlm·km$^{-2}$) times a factor $\mathcal{H}(\boldsymbol{\beta})$ that depends on the nadir observation angle, $z$, the effects of the obstacles, the spectral composition of the mix of artificial light sources, their angular emission patterns, and atmospheric conditions, collectively denoted by the parameter vector **β**. The factor

$$\mathcal{H}(\boldsymbol{\beta}) = K_{DNB}\, Q(z) \int_{\lambda=0}^{\infty} T_{DNB}(\lambda)\, T_{atm}(\lambda,z)\, \frac{\rho(\lambda)}{\pi} \sum_{i=1}^{I} \gamma_i\, \widehat{\Phi}_i(\lambda) \left(\frac{\lambda}{hc}\right) d\lambda \quad (12)$$

can be interpreted as the radiance in W·m$^{-2}$·sr$^{-1}$ reported by the VIIRS-DNB per emitted lm·m$^{-2}$, and has units W·sr$^{-1}$·lm$^{-1}$.

After an outdoor lighting remodeling, the radiance reported by the VIIRS-DNB may vary due to the variation of the density of emitted luminous flux, $N$, and/or to the variation of the composition of the sources mix $\mathcal{H}(\boldsymbol{\beta})$, through the subset $\{\gamma_i, \widehat{\Phi}_i\,;\, i=1,\dots,I\}$ of the **β** parameter vector.

*2.4 Output indicators and ratios*

In order to analyze lighting remodeling processes it is convenient to express the satellite measurements in terms of the lumen density emitted by the installations that do not change ($N_0$), and the ones emitted by the installations modified in the process, before ($N_b$) and after ($N_a$) the change of lamps and luminaires. The ratio $R_{ab}$ of the modelled VIIRS-DNB measurements after/before the change is

$$R_{ab} = \frac{L_{total,a}}{L_{total,b}} = \frac{N_0\, \mathcal{H}(\boldsymbol{\beta}_0) + N_a\, \mathcal{H}(\boldsymbol{\beta}_a)}{N_0\, \mathcal{H}(\boldsymbol{\beta}_0) + N_b\, \mathcal{H}(\boldsymbol{\beta}_b)} \quad (13)$$

where for the sake of simplicity it is assumed that the average atmospheric conditions and ground albedo are the same in the periods immediately before and after the remodeling, such that the differences in the parameter vectors $\boldsymbol{\beta}_0$, $\boldsymbol{\beta}_a$, and $\boldsymbol{\beta}_b$ are only due to the different spectral and geometric properties of the sources and luminaires. From the ratio $R_{ab}$, the known values of $N_a$ and $N_b$, and the calculated values of $\mathcal{H}$ it is possible to estimate the amount of emitted lumen from the installations that did not undergo modifications, as:

$$N_0 = \frac{R_{ab} N_b\, \mathcal{H}(\boldsymbol{\beta}_b) - N_a\, \mathcal{H}(\boldsymbol{\beta}_a)}{\mathcal{H}(\boldsymbol{\beta}_0)[1 - R_{ab}]} \quad (14)$$

And, subsequently, the fractions of emitted lumen of the public outdoor lighting system before ($\Gamma_b$) and after ($\Gamma_a$) the remodeling, vs the total number of emitted lumen, are

$$\Gamma_i = \frac{N_i}{N_0 + N_i} = \frac{1}{1 + (N_0/N_i)} \qquad (15)$$

where $i \in \{b, a\}$.

## 3. Results

*3.1. The Ribeira municipality lighting transformation*

Ribeira (42°33'23"N, 8°59'32" W) is a municipality of 27 000 inhabitants and 68.83 km² belonging to Galicia, autonomous community of the kingdom of Spain. It is located on the western Galician Atlantic coast and it is one of the main in-shore and high-seas fishing harbors of the community. In the year 2015 the municipality underwent a full renovation of its public streetlight system, substituting a total of 8395 high-pressure sodium, metal halide and mercury vapor street lamps by the same amount of 4000 K LED sources. In this operation, a total of 93.2x10$^6$ installed lumen were replaced by 28.7 x10$^6$ lumen of the new lamps. This strong reduction on the installed lumen was made possible by an improved luminaire efficiency, a larger utilance (and hence a larger overall utilization factor), and the reduction of the former street illuminance levels to accommodate them to the (still high) maximum values established in current regulations. The main characteristics of the lamp inventory before and after the transformation are summarized in Table 1.

The actual number of lm exiting the luminaires is generally smaller than the maximum amount of light emitted by the lamps at the beginning of their useful life, due to lamp aging (accounted for by the depreciation factor) and to absorption and light screening within the luminaire enclosure (accounted for by the luminaire efficiency factor). The legacy system of the Ribeira municipality used traditional luminaires for which the average efficiency factor is of order $g_i$ = 0.7. The lamps were substituted in 2015, and an average lumen depreciation factor $f_i$ = 0.9 can be assigned to them based on a nominal maintenance factor of 0.8. The new luminaires, mostly based on flat LED boards, have an efficiency close to 1.0. On the other hand, since the VIIRS-DNB measurements used for monitoring this change correspond to the year 2017 (two years after the change to long-life LED lamps) we adopted a lumen depreciation factor of order 1.0 for the new installation at the time of measurement. These factors translate into an effective maximum flux leaving the luminaires of 58.7x10⁶ and 28.7x10⁶ lm before and after the change, respectively.

**Table 1.** Lumen emission budget, before and after the Ribeira municipality transformation.

| **Installed lumen (initial)** | | Num Lamps: 8395 | **Installed lumen (final)** | | Num Lamps: 8395 |
|---|---|---|---|---|---|
| Lamp type | % over total num. lamps | Lumen per lamp $N_{Li}$ | Lamp type | % over total num. lamps | Lumen per lamp $N_{Li}$ |
| HPS 070 | 40 | 6 600 | LED 020 | 30 | 1 600 |
| HPS 100 | 25 | 10 500 | LED 040 | 35 | 3 200 |
| HPS 150 | 10 | 16 500 | LED 050 | 10 | 4 000 |
| MH 150 | 10 | 17 000 | LED 060 | 5 | 5 100 |
| MV 125 | 10 | 6 200 | LED 020b | 5 | 2 200 |
| MH 450 | 5 | 37 350 | LED 024 | 5 | 1 960 |
| | | | LED 012 | 5 | 1 200 |
| | | | LED 110 | 5 | 17 820 |
| Total installed (lm) initial | | 93 205 488 | Total installed (lm) final | | 28 660 530 |
| Depreciation factor $f_i$ = 0.9 Luminaire efficiency $g_i$ = 0.7 | | | Depreciation factor $f_i$ = 1.0 Luminaire efficiency $g_i$ = 1.0 | | |
| Total emitted by luminaires (lm) | | 58 719 457 | Total emitted by luminaires (lm) | | 28 660 530 |
| ULOR $u_i$ = 0.1 Regulation factor $r_i$ = 1.0 | | | ULOR $u_i$ = 0.0 Regulation factor $r_i$ = 0.5 | | |
| Total regulated on surfaces (lm) | | 52 847 511 | Total regulated on surfaces (lm) | | 14 330 265 |
| Average flux density on surfaces $F_S/S$ ($Mlm \cdot km^{-2}$) = $N_b$ | | 0.768 | Average flux density on surfaces $F_S/S$ ($Mlm \cdot km^{-2}$) = $N_a$ | | 0.208 |

The fraction of light emitted by the luminaires directly into the atmosphere, without interacting with neighboring surfaces, can be estimated to be of order $u_i$=0.1 for the legacy lighting system. This is due to the difficulty of efficaciously collimating the light emitted by the relatively large active emitting surface of gas discharge lamps, the use of curved glass enclosures and the traditional practice of mounting the luminaires tilted by some degrees above the horizontal. These factors are much less influential in LED based luminaires with flat printed circuit boards. We have consequently adopted an approximate factor $u_i$=0.0 for the new installation. Regarding intensity reductions at the central part of the night, the legacy system operated full time at the nominal maximum intensity, so that $r_i$=1.0 before the changes. The new system is fitted with regulating drivers that reduce the light emissions to $r_i$=0.5 times the maximum value, between 00:00h and 07:00h (geographic time zone UTC–1, standard time CET=UTC+1, savings time CEST=UTC+2), time interval that includes the instant when the VIIRS-DNB acquires the nightly radiance readings, about 01:30 solar mean local time. The total amount of light from the streetlights shed on the surfaces of the municipality at this time of the night was then 52.8x10$^6$ and 14.3x10$^6$ lm before and after the change, respectively. The average flux density produced by this system on the surface of the municipality (i.e. the average irradiance) is 0.768 and 0.208 Mlm·km$^{-2}$, respectively.

*3.2. Modeled evolution of the VIIRS-DNB recorded radiance*

The modelled evolution of the VIIRS-DNB recorded radiance, $L(z)$, can be computed using Eq.(11). The data in the above subsections provide the values of the input variables $F_S/S$, $K_{DNB}$, and $\gamma_i$, the latter being deduced from Eq.(3) from the known values of $F_S$, $n_i$, $f_i$, $g_i$, $r_i$, $u_i$, and $N_{Li}$. The remaining functions and parameters required for predicting the detected radiance are the lamp spectra $\widehat{\Phi}_i(\lambda)$, the spectral reflectance of the surfaces $\rho(\lambda)$, the spectral attenuation of the atmosphere $T_{atm}(\lambda, z)$, the spectral sensitivity function of the detector $T_{DNB}(\lambda)$, and the shadowing transmittance function $Q(z)$.

The unnormalized lamp spectral power distributions, $\Phi_i(\lambda)$, from which the ones normalized to 1 lm can be calculated using Eq.(1), can be obtained from the manufacturers or downloaded from online databases [27-28]. The ground spectral reflectance $\rho(\lambda)$ to be used for this calculation is the one of the materials located immediately beneath or next to the streetlights, usually a mix of different types of concrete and asphalt. For the purposes of this work we adopted a combination 1:0.25:0.25 of asphalt and two types of concrete from the USGS Spectral Library Version 7 database [43] whose resulting spectral reflectance $\rho(\lambda) = \rho_{ground}(\lambda)$ is displayed in Figure 1.

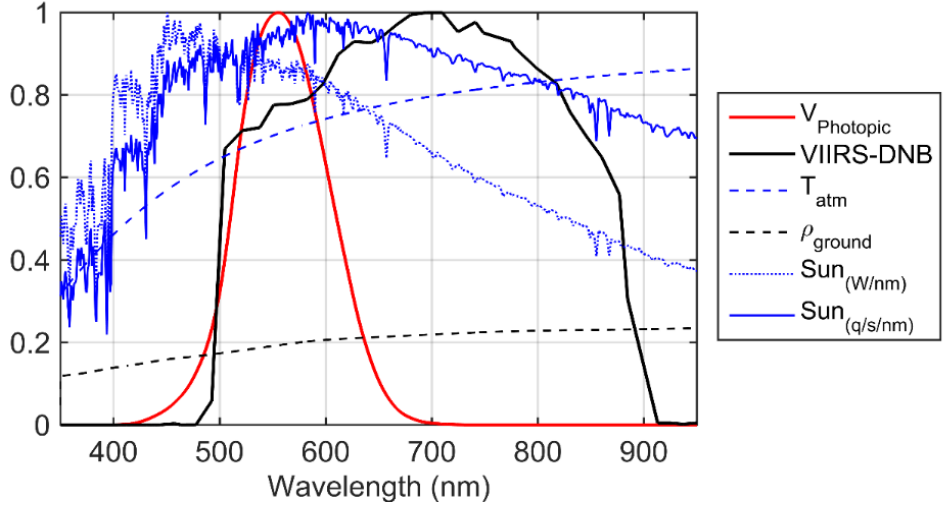

**Figure 1:** Photopic spectral sensitivity function of the human visual system $V(\lambda)$ (full line, red), spectral passband of the VIIRS-DNB radiometer $T_{DNB}(\lambda)$ (full line, black), atmospheric transmittance function averaged over zenith angles $T_{atm}(\lambda)$ (dashed line, blue, see text for details), ground reflectance $\rho(\lambda)$ (dashed line, black, see text for details), extra-atmospheric Sun spectral power distribution $E_{Sun}(\lambda)$ in W·nm$^{-1}$ normalized to 1 at its maximum (dotted line, blue), and extra-atmospheric Sun spectral power distribution $E_{Sun}(\lambda)\left(\frac{\lambda}{hc}\right)$ in light quanta/s/nm normalized to 1 at its maximum (full line, blue). Color online.

The spectral attenuation of the atmosphere was calculated for typical visibility conditions with molecular (Rayleigh) and aerosol optical depths given by Eqs.(5) and (6), respectively with an AOD $\tau_{A,0}(\lambda_0)$=0.2 at the reference wavelength $\lambda_0$=500 nm, and Ångström exponent $\alpha = 1$. The value of $T_{atm}(\lambda, z)$ was averaged over a uniform probability density distribution of zenith angles $z \in [-60°, 60°]$, consistent with the expected distribution of the VIIRS-DNB measurement nadir angles over a whole year for this latitude [35]. The resulting $T_{atm}(\lambda)$ function is displayed in Fig. 1.

The combination of the spectral distributions of the radiant fluxes emitted by the 8395 luminaires, per unit area of the municipality, gives rise to the average spectral irradiances $E_S^-(\lambda)$ incident on the ground before and after remodeling, whose shapes are depicted in Figure 2 as $E_{ground,b}^-(\lambda)$ and $E_{ground,a}^-(\lambda)$, respectively. The plots in the upper panel of this figure correspond to the spectral irradiances in W·m$^{-2}$·nm$^{-1}$ and the ones in the lower panel to the spectral irradiances in photons·s$^{-1}$·m$^{-2}$·nm$^{-1}$, both normalized to 1 at their maxima for display purposes. After reflection off the ground and propagation through the atmosphere in the conditions described in the previous paragraph, the radiances at the top of the atmosphere, $L^+_{TOA}(\lambda, z)$, given by Equation (7), averaged over zenith angles $z \in [-60°, 60°]$ and with the obstacle transmittance function $Q(z)$ provisorily set for convenience equal to 1 (see subsection 3.4 for an estimate of its actual value) are plotted in Fig. 2 as $L^+_{TOA,b}$ and $L^+_{TOA,a}$, corresponding to the situations before and after the change, respectively. Again, all plots are normalized to 1 at their maxima, being the upper panel corresponding to spectral radiances in energy units W·m$^{-2}$·sr$^{-1}$·nm$^{-1}$ and the lower one in photon number units photons·s$^{-1}$·m$^{-2}$·sr$^{-1}$·nm$^{-1}$. The spectral passband of the VIIRS-DNB radiometer $T_{DNB}(\lambda)$ is also shown for reference.

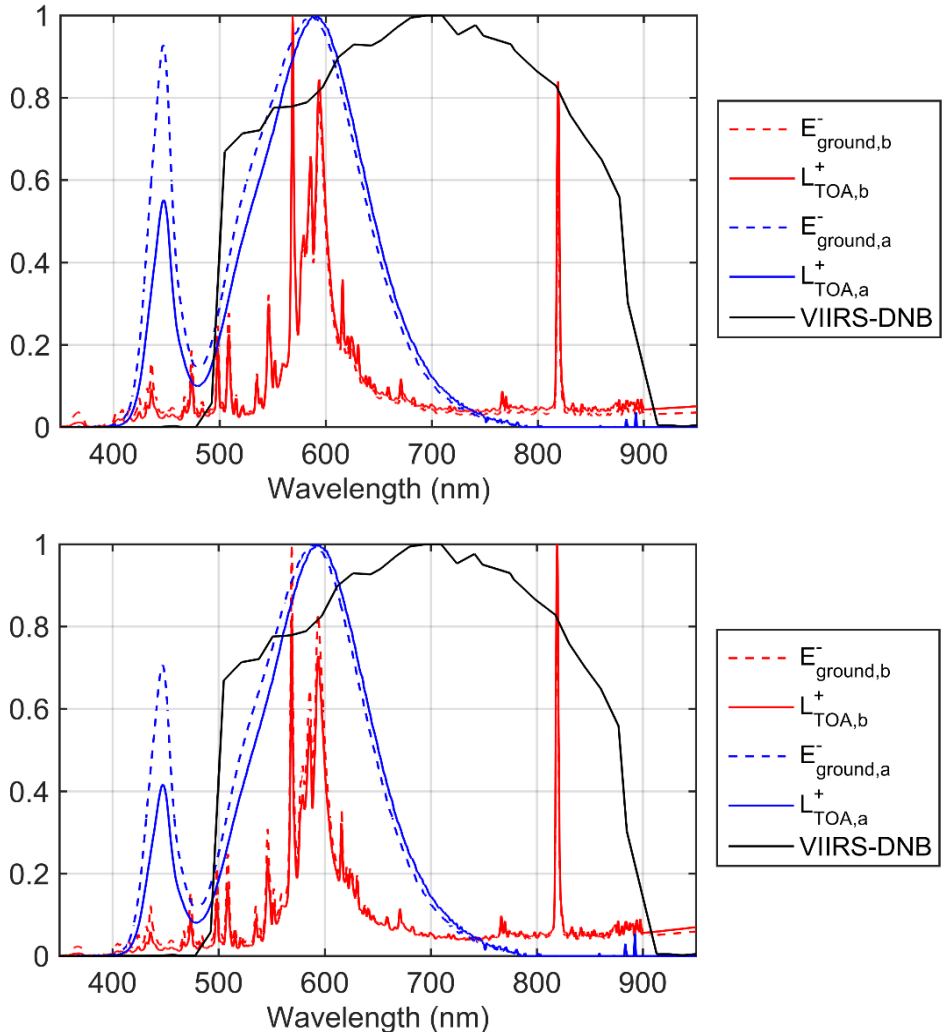

**Figure 2:** Spectral irradiance incident on the ground before and after the remodeling, $E^-_{ground,b}(\lambda)$ and $E^-_{ground,a}(\lambda)$, respectively and the corresponding top of atmosphere spectral radiances $L^+_{TOA,b}$ and $L^+_{TOA,a}$. Upper panel in energy units W·m$^{-2}$·nm$^{-1}$ for the irradiance and W·m$^{-2}$·sr$^{-1}$·nm$^{-1}$ for the radiance;

Lower panel in units photons·s$^{-1}$·m$^{-2}$·nm$^{-1}$ and photons·s$^{-1}$·m$^{-2}$·sr$^{-1}$·nm$^{-1}$, respectively. All plots normalized to 1 at their maxima. The VIIRS-DNB spectral passband $T_{DNB}(\lambda)$ is also shown.

These data may be used in combination with Eqs.(11) and (12) to obtain the predicted VIIRS-DNB radiances produced by the public lighting system $L_b = N_b\,\mathcal{H}(\boldsymbol{\beta}_b) = 8.2186$ nW·cm$^{-2}$·sr$^{-1}$ and $L_a = N_a\,\mathcal{H}(\boldsymbol{\beta}_a) = 1.6897$ nW·cm$^{-2}$·sr$^{-1}$, in values averaged over the year and over the municipality territory for the scenarios before and after the change, respectively. The associated $\mathcal{H}(\boldsymbol{\beta})$ functions have the values $\mathcal{H}(\boldsymbol{\beta}_b) = 0.1070 \times 10^{-3}$ and $\mathcal{H}(\boldsymbol{\beta}_a) = 0.0812 \times 10^{-3}$, in W·sr$^{-1}$·lm$^{-1}$. Recall that $\mathcal{H}(\boldsymbol{\beta})$ can be interpreted as the radiance in W·m$^{-2}$·sr$^{-1}$ reported by the VIIRS-DNB per emitted lm·m$^{-2}$, so if the average emissions within a VIIRS-DNB pixel would amount to one lm·m$^{-2}$ (or, equivalently, one Mlm·km$^{-2}$) and not be blocked by obstacles, the reported VIIRS-DNB radiance readings, under the observing conditions used for this calculation, would be 10.70 and 8.12 nW·cm$^{-2}$·sr$^{-1}$, respectively.

Regarding the effect of using the solar-calibrated $K_{DNB}$ constant in Eq.(11) of section 2.3 for reporting the radiance in power units from the photon·s$^{-1}$ counts when the spectrum of the incident radiance does not match exactly the solar one, we found that for the light emitted by the public streetlight system the reported VIIRS-DNB radiance readings are expected to be 0.9571 and 0.8946 times the actual values before and after the changes, respectively (this effect is already taken into account in the $\mathcal{H}(\boldsymbol{\beta})$ definition and hence compensated for in the calculation of the output indicators).

*3.3. Observed evolution of the VIIRS-DNB recorded radiance*

To evaluate the changes in the TOA radiance recorded by the radiometer aboard the Suomi-NPP satellite we used the Annual VNL V2 VIIRS-DNB average-masked composite series [20, 40] analyzing the evolution from 2014 to 2020, both years included. The average radiance within the municipality term of Ribeira was extracted with QGIS using the official shapefiles of the municipal, provincial and autonomous areas and municipal, provincial and Autonomous Community boundaries registered in the Spanish Central Register of Cartography [44]. All operations were performed in the native EPSG:4326 coordinate reference system of the VIIRS-DNB tiff layers to avoid potential bias derived from interpolated reprojections.

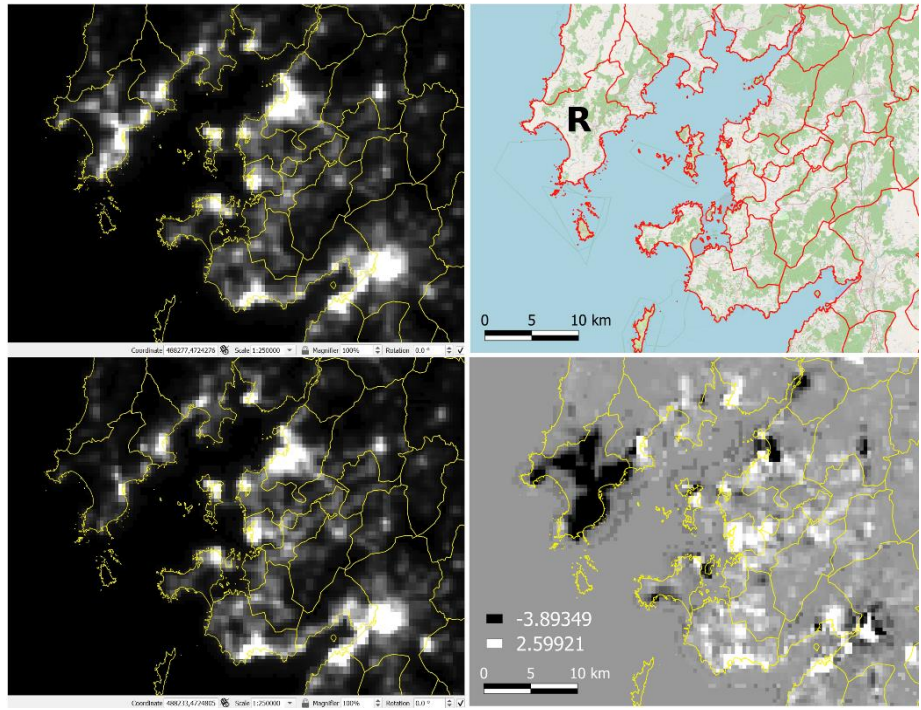

**Figure 3.** Left column: Annual VNL V2 VIIRS-DNB average-masked composites of 2014 (top) and 2017 (bottom), in the Ribeira municipality area. Right column: Geographic map showing the limits of Ribeira (R) and surrounding municipalities (top) and map of the radiance differences 2017–2014 (bottom). Grayscale in nW·cm$^{-2}$·sr$^{-1}$. Credits: for VIIRS-DNB images, Elvidge et al. [20] and Earth Observation Group [40], for municipality limits BDLJE 2012 CC-BY 4.0 ign.es, for map OpenStreetMap.

Figure 3 displays the composites of the years 2014 (left column, top) and 2017 (left, bottom), the geographical map of Ribeira and surrounding municipalities (right column, top), and the radiance differences 2017–2014 (right, bottom).

The evolution of the measured radiance of the Ribeira municipality, averaged over time and also over its territory, is shown in Figure 4. The square symbols show the annual values deduced from the composites *VNL_v2_npp_YYYY_global_vcmslcfg_cYYYYMMDD0000.average_masked.tif*. The dots are monthly averaged values, as reported by the Radiance Light Trends app [45]. The ticks in the horizontal axis correspond to January 1st of the years indicated in the labels. The yearly average symbols (squares) are located in the middle of the corresponding year. The average annual radiances from 2014 to 2020 deduced from the VIIRS-DNB composites are 7.00, 6.08, 3.18, 2.58, 2.38, 2.46 and 2.37 nW·cm$^{-2}$·sr$^{-1}$, respectively.

Figure 4 shows the sharp decline of the detected radiance that took place along the year 2015, when the remodeling of the public outdoor lighting system was carried out. The monthly averages in 2014 and before show a large variability, substantially reduced from 2016 onwards, an effect possibly related to the flicker reduction reported in other gas-discharge to LED lamp transitions [46]. For our calculations in the next section we took the measured values (averaged over the corresponding year and over the municipality territory) $L_{DNB,b} = 7.00$ nW·cm$^{-2}$·sr$^{-1}$ of the year 2014 as representative of the situation before the changes, and the $L_{DNB,a} = 2.58$ nW·cm$^{-2}$·sr$^{-1}$ value of the year 2017 as representative of the situation after the changes. The value of 2016 (when the new lighting system was already operative), slightly larger than the one of 2017 and following years, was not used for the analysis due to the presence of a small but conspicuous subset of very bright pixels within Ribeira and neighboring municipalities, located in an unpopulated rural area without roadways, buildings or equipment, only appearing in the composite of that year and whose origin could not be identified by the municipality technical services, being possibly an uncorrected artifact.

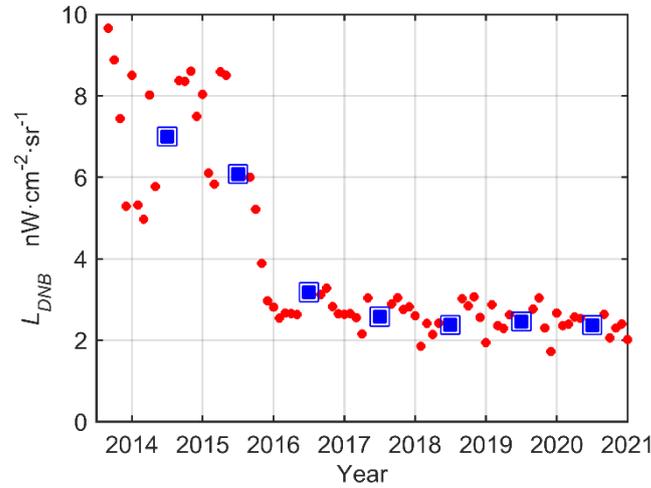

**Figure 4.** Temporal evolution of the VIIRS-NPP radiance readings, spatially averaged over the territory of the Ribeira municipality, $L_{DNB}$. Squares: Average-masked yearly values. Dots: Monthly averages. The ticks in the horizontal axis correspond to January 1st of the years indicated in the labels. The yearly average symbols (squares) are located in the middle of the corresponding year. See text for details.

*3.4. Output indicators*

The spatial lumen densities $N_b$ and $N_a$ derived from the public lighting inventory (subsection 3.1), in combination with the model predictions for $L_b$, $L_a$, $\mathcal{H}(\boldsymbol{\beta}_b)$, and $\mathcal{H}(\boldsymbol{\beta}_a)$ obtained in section 2.2., and the actual readings $L_{DNB,b}$, $L_{DNB,a}$ of the VIIRS-DNB for the situations before and after the remodeling described in subsection 3.3. allow to estimate the lumen density emissions $N_0$ of the sources not belonging to the public system, by using Eq.(14), and the proportion of emissions of the public system versus total emissions before ($\Gamma_b$) and after ($\Gamma_a$) the remodelling, by using Eq.(15). Some information can also be obtained about the obstacle transmittance function $Q(z)$ that was arbitrarily set to 1 in the calculations of section 3.2.

In order to apply Eqs.(13)-(15) it is necessary to estimate the value of $\mathcal{H}(\boldsymbol{\beta}_0)$, the function relating the lumen density emissions that do not belong to the public outdoor lighting system to the radiance detected by the VIIRS-DNB from this subset of sources. An important input is the spectral composition of the sources. There is presently no detailed information about the composition of this lamp subset mix, but the fact that most of these sources belong to large fishing harbor installations and industrial zones illuminated by "cold" white light of typical CCTs of order 4000 K makes it sensible to adopt for $\mathcal{H}(\boldsymbol{\beta}_0)$ the same value as for $\mathcal{H}(\boldsymbol{\beta}_a)$. The value of $\mathcal{H}(\boldsymbol{\beta}_0)$ is not modified throughout the transformation process. Other factors that determine the value of $\mathcal{H}(\boldsymbol{\beta}_0)$, such as obstacle transmittances and atmospheric conditions, may be assumed to be similar to the ones in the remaining $\mathcal{H}$ functions, as a reasonable first approximation.

Recall that any constant scaling factor present in the three $\mathcal{H}$ functions cancels out in Eqs. (13)-(15) and does not affect the estimation of $N_0$, $\Gamma_b$ and $\Gamma_a$. This applies in particular to the actual value of the obstacle transmittance term $Q(z)$ averaged over zenith angles that in subsection 2.2 was set for convenience equal to 1. Accordingly, we obtain the estimates $N_0 = 0.2578$ Mlm·km$^{-2}$, for the lumen density over surfaces due to sources different from the public outdoor streetlight system, $\Gamma_b = 74.86\%$ for the percent contribution of the public streetlights to the total lumen emitted in the territory before remodeling and $\Gamma_a = 44.68\%$ for their percent contribution after remodeling.

The results of the previous paragraph are the main outcomes of this application example. Note however that some complementary information can be obtained from them, namely the estimated value of any global constant missing in the value of the $\mathcal{H}$ functions. To that end, recall that the predicted values for the VIIRS-DNB total detected signal are, according to the numerator and denominator of Eq.(13) $L_{total,b} = N_0\,\mathcal{H}(\boldsymbol{\beta}_0) + N_b\,\mathcal{H}(\boldsymbol{\beta}_b) = 10.31$ and $L_{total,a} = N_0\,\mathcal{H}(\boldsymbol{\beta}_0) + N_a\,\mathcal{H}(\boldsymbol{\beta}_a) = 3.78$, both in nW·cm$^{-2}$·sr$^{-1}$. The absolute on-orbit measured values are however $L_{DNB,b} = 7.00$ and $L_{DNB,a} = 2.58$ nW·cm$^{-2}$·sr$^{-1}$. The corresponding ratios measured-to-modelled are then $L_{DNB,b}/L_{total,b} \cong L_{DNB,a}/L_{total,a} = 0.68$. This suggests that the product of all constant factors adopted in our model example should be multiplied by this ratio to match the actual absolute measurements. This allows to estimate that the zenith-averaged $Q(z)$ obstacle transmittance should have a value of a 68%, instead of the 100% adopted for convenience in the model calculation of subsection 2.2. Note however that the estimated values assigned to other parameters besides $Q(z)$ are also approximate, and that the 0.68 matching ratio applies to their overall product, so all individual parameters can contribute to a bigger or lesser extent to this factor.

## 4. Discussion

The core of the approach described in this paper, basically contained in Eqs.(13)-(15), relies on the use of $\mathcal{H}(\boldsymbol{\beta})$ functions connecting ground emissions to satellite radiance readings. The particular implementation of the model for calculating the $\mathcal{H}(\boldsymbol{\beta})$ function described in subsections 2.1 to 2.3 could be replaced, when convenient, by models of bigger complexity in function of the situations to be analyzed. Note that the $L(z)$ radiance calculated in Eq.(11) is linear on the product of the relevant parameters of the ground lighting installation, and the uncertainties in the values of these parameters will propagate to the final results according to the usual rules.

The results presented in section 3 are intended as a practical example of application of the method described in this paper and are subjected to several sources of uncertainty. One of them is the actual spectral reflectance distribution $\rho(\lambda)$ of the surfaces illuminated by the artificial lights. In an extended municipality with a main town of small size, many dispersed villages, and a large extension of woods and agricultural fields of small farms, most of the artificial light is reflected on the pavement of the streets or roadways immediately beneath the light poles, where the light spots have non-negligible illuminance. These asphalt and concrete artificial surfaces only cover a small fraction of the surface of the pixels of commonly available ground reflectance products (e.g. MODIS) whose spectral reflectance in areas like this municipality are mostly determined by the reflectance of the unlit vegetal cover in agricultural fields or by a combination of the reflectance of the streets, façades and roofs in urban sites. The traditional roof covers in Western Galicia (clay tiles) have a spectral reflectance substantially different from that of the streets (asphalt and concrete). This makes the spatially averaged reflectance values derived from usual satellite products not representative of the actual reflectance of the illuminated surfaces.

Other relevant sources of uncertainty are those related with the particular state of the atmosphere in each of the individual VIIRS-DNB passes from which the yearly composites are built. We have adopted in this work usual values for the AOD in this region of the Galician community, but a finer analysis should use specific AOD nighttime values estimated for each individual pass from the daytime values immediately previous and posterior to the corresponding night. The same individual estimation should be applied to the airmass number, as a function of the actual nadir angle of each satellite pass. Furthermore, when substituting the annual average of the product of various factors by the product of their averages some bias is inevitably introduced if some of these factors are partially correlated, as it arguably happens with the obstacle transmittance and the airmasses, both expected to be higher for larger zenith angles. All these uncertainty factors, taken together, suggest the desirability of exercising some degree of prudence in the interpretation of these (and other related works) results regarding the actual values of some parameters.

The ratios $\Gamma_b$ and $\Gamma_a$ calculated with the methodology described in this work, however, are relatively robust against these uncertainties, since several uncertain multiplicative factors cancel out when taking the radiance ratios. This allows us to confidently estimate that for the case example studied in this work the public streetlight system was responsible of the 74.86% of the light pollution emissions (at the time of the night when the VIIRS-DNB radiometer took its readings) before the remodeling, dropping to a still significant 44.68% after the transformation.

Eqs. (13)-(15) can be used to analyze the outcomes of other outdoor lighting remodeling operations, for which detailed information exists in the literature. As an example, they can be applied to the lighting reduction emissions in the small rural Catalonian nuclei Deltebre and Riumar, carried out in 2013 and reported in Estrada-García et al [22]. The emissions in klm and the VIIRS-DNB readings are described in tables 3 and 4 of that paper. Since the transformation consisted on replacing large power high-pressure sodium lamps by lower power versions of the same sources, $\mathcal{H}(\boldsymbol{\beta}_0)$, $\mathcal{H}(\boldsymbol{\beta}_b)$, and $\mathcal{H}(\boldsymbol{\beta}_a)$ are equal to each other and cancel out in Eqs.(13) and (14), so the $\Gamma_b$ and $\Gamma_a$ ratios can be directly calculated from the ground-level emissions. It is found that the public outdoor lighting systems in Deltebre accounted for 89.9% of the total emissions before remodeling, decreasing to 79.4% after the public works. In the smaller Riumar nuclei, the values were 99.5% and 98.6%, respectively, consistent with the fact that practically all outdoor lighting in that place belonged to the public streetlight system. These results are expectedly different from the ones obtained in places where the publicly owned streetlights make a small part of the total light emissions, as e.g. the case of Tucson (AZ, USA) reported by Kyba et al. [23]. These results also highlight the importance of carrying out emission estimations adapted to the particular regions of the world under study, expecting that the situation will largely vary from one to another depending on the particular importance that publicly owned streetlight systems may have in each of them.

5. Conclusions

We describe in this work a practicable method for obtaining information about the artificial light emissions at the municipality level by combining information from public inventories of installed lamps (and their changes in time) and radiance measurements obtained from low Earth orbit radiometers. The method has been particularized here for the spectral passband of the VIIRS-DNB radiometer onboard the Suomi-NPP satellite, but its extension to other radiometer bands is immediate. The application of this approach to a case study of large lighting remodeling at the municipality of Ribeira (Galicia, Spain, European Union) shows that the public streetlight system accounted for a 74.86% of the total light pollution emissions before the remodeling, decreasing to a 44.68% after the changes were made. These results highlight the importance (both for scientific purposes and for territorial planning) of making specific estimations to ascertain the contribution of the public streetlight system to the overall light pollution source, as well as the convenience of analyzing the contributions of other kind of sources, especially in those territories where the public streetlights already reduced considerably their weight.

**Acknowledgements**

Thanks are due to the municipality of Ribeira, in Galicia (Spain), and Ferrovial, for their collaboration in the realization of this study.

**Funding sources**

CB acknowledges funding from Xunta de Galicia/FEDER, grant ED431B 2020/29.

**References**

[1] Longcore T, Rich C. Ecological light pollution. Frontiers in Ecology and the Environment 2004;2:191-198.

[2] Hölker F, Wolter C, Perkin EK, Tockner K. Light pollution as a biodiversity threat. Trends in Ecology and Evolution 2010;25:681-682

[3] Gaston KJ, Bennie J, Davies TW, Hopkins J. The ecological impacts of nighttime light pollution: a mechanistic appraisal. Biological Reviews 2013;88:912–927

[4] Pun CSJ, So CW. Night-sky brightness monitoring in Hong Kong. Environmental Monitoring and Assessment 2012;184(4): 2537-2557. https://doi.org/10.1007/s10661-011-2136-1

[5] Bará S. Anthropogenic disruption of the night sky darkness in urban and rural areas. Royal Society Open Science 2016;3:160541. https://doi.org/10.1098/rsos.160541

[6] Posch T, Binder F, Puschnig J. Systematic measurements of the night sky brightness at 26 locations in Eastern Austria. Journal of Quantitative Spectroscopy & Radiative Transfer 2018;211:144–165. https://doi.org/10.1016/j.jqsrt.2018.03.010

[7] Bertolo A, Binotto R, Ortolani S, Sapienza S. Measurements of Night Sky Brightness in the Veneto Region of Italy: Sky Quality Meter Network Results and Differential Photometry by Digital Single Lens Reflex. J. Imaging 2019;5:56. https://doi.org/10.3390/jimaging5050056

[8] Hänel A, Posch T, Ribas SJ, Aubé M, Duriscoe D, Jechow A, Kollath Z, Lolkema DE, Moore C, Schmidt N, Spoelstra H, Wuchterl G, Kyba CCM. Measuring night sky brightness: methods and challenges. Journal of Quantitative Spectroscopy & Radiative Transfer 2018;205:278–290. https://doi.org/10.1016/j.jqsrt.2017.09.008

[9] Kuechly HU, Kyba CCM, Ruhtz T, Lindemann C, Wolter C, Fischer J, Hölker F. Aerial survey and spatial analysis of sources of light pollution in Berlin, Germany. Remote Sensing of Environment 2012;126:39–50. https://doi.org/10.1016/j.rse.2012.08.008

[10] Fiorentin P, Bettanini C, Bogoni D. Calibration of an Autonomous Instrument for Monitoring Light Pollution from Drones, Sensors 2019;19:5091. https://doi.org/10.3390/s19235091

[11] Bouroussis CA, Topalis FV. Assessment of outdoor lighting installations and their impact on light pollution using unmanned aircraft systems - The concept of the drone-goniophotometer. Journal of Quantitative Spectroscopy & Radiative Transfer 2020;259:107155 https://doi.org/10.1016/j.jqsrt.2020.107155

[12] Gyuk G, Garcia J, Tarr C, Walczak K. Light Pollution Mapping from a Stratospheric High-Altitude Balloon Platform. International Journal of Sustainable Lighting 2021:23(1):20-32. https://doi.org/10.26607/ijsl.v23i1.106

[13] Kyba CCM, Garz S, Kuechly H, Sánchez de Miguel A, Zamorano J, Fischer J, Hölker F. High-Resolution Imagery of Earth at Night: New Sources, Opportunities and Challenges, Remote Sens. 2015;7:1-23. https://doi.org/10.3390/rs70100001

[14] Elvidge CD, Baugh K, Zhizhin M, Hsu FC, Ghosh T. VIIRS night-time lights. International Journal of Remote Sensing 2017;38(21):5860-5879 https://doi.org 10.1080/01431161.2017.1342050

[15] Li X, Elvidge C, Zhou Y, Cao C, Warner T. Remote sensing of night-time light. International Journal of Remote Sensing, 2017;38(21):5855-5859. https://doi.org/10.1080/01431161.2017.1351784

[16] Stefanov WL, Evans CA, Runco SK, Wilkinson MJ, Higgins MD, Willis K. Astronaut Photography: Handheld Camera Imagery from Low Earth Orbit, in J.N. Pelton et al. (eds.), Handbook of Satellite Applications, Switzerland:Springer International Publishing;2017. https://doi.org/10.1007/978-3-319-23386-4_39

[17] Román MO, Wang Z, Sun Q, Kalb V, Miller SD, Molthan A, Schultz L, Bell J, Stokes EC, Pandey B, Seto KC, Hall D, Oda T, et al. NASA's Black Marble nighttime lights product suite, Remote Sensing of Environment 2018;210:113-143. https://doi.org/10.1016/j.rse.2018.03.017

[18] Sánchez de Miguel A, Christopher C.M. Kyba, Martin Aubé, Jaime Zamorano, Nicolas Cardiel, Carlos Tapia, Jon Bennie, Kevin J. Gaston. Colour remote sensing of the impact of artificial light at night (I): The potential of the International Space Station and other DSLR-based platforms. Remote Sensing of Environment 2019;224:92–103. https://doi.org/10.1016/j.rse.2019.01.035

[19] Levin N, Kyba CCM, Zhang Q, Sánchez de Miguel A, Román MO, LiL, Portnov BA, Molthan AL, Jechow A, Miller SD, Wang Z, Shrestha RM, Elvidge CD. Remote sensing of night lights: A review and an outlook for the future. Remote Sensing of Environment 2020;237:111443. https://doi.org/10.1016/j.rse.2019.111443

[20] Elvidge CD, Zhizhin M, Ghosh T, Hsu FC, Taneja J. Annual time series of global VIIRS nighttime lights derived from monthly averages: 2012 to 2019. Remote Sensing 2021;(5) 922, https://doi.org/10.3390/rs13050922

[21] Sánchez de Miguel A, Zamorano J, Aubé M, Bennie J, Gallego J, Ocaña F, Pettit DR, Stefanov WL, Gaston KJ. Colour remote sensing of the impact of artificial light at night (II): Calibration of DSLR-based images from the International Space Station. Remote Sensing of Environment, 2021:112611. https://doi.org/10.1016/j.rse.2021.112611.

[22] Estrada-García R, García-Gil M, Acosta L, Bará S, Sanchez de Miguel A, Zamorano J. Statistical modelling and satellite monitoring of upward light from public lighting. Lighting Research and Technology 2016;48:810-822. https://doi.org/10.1177/1477153515583181

[23] Kyba C, Ruby A, Kuechly H, et al. Direct measurement of the contribution of street lighting to satellite observations of nighttime light emissions from urban areas. Lighting Research & Technology, 2021;53(3):189-211. https://doi.org/10.1177/1477153520958463

[24] Bará S, Rodríguez-Arós A, Pérez M, Tosar B, Lima RC, Sánchez de Miguel A, Zamorano J. Estimating the relative contribution of streetlights, vehicles, and residential lighting to the urban night sky brightness. Lighting Res. Technol. 2019;51:1092–1107. https://doi.org/10.1177/1477153518808337

[25] Barentine JC, Kundracik F, Kocifaj M, Sanders JC, Esquerdo GA, Dalton AM, Foott B, Grauer A, Tucker S, Kyba CCM. Recovering the city street lighting fraction from skyglow measurements in a large-scale municipal dimming experiment. Journal of Quantitative Spectroscopy & Radiative Transfer 2020;253:107120 https://doi.org/10.1016/j.jqsrt.2020.107120

[26] Sánchez de Miguel A, Aubé M, Zamorano J, Kocifaj M, Roby J, Tapia C. Sky Quality Meter measurements in a colour-changing world. Monthly Notices of the Royal Astronomical Society 2017;467(3):2966-2979. https://doi.org/10.1093/mnras/stx145

[27] Tapia C, Sánchez de Miguel A, Zamorano, J. 2017. LICA-UCM Lamps Spectral Database 2.6.. https://eprints.ucm.es/id/eprint/40841/1/LICA_Spectra_database_v2_6.pdf

[28] Roby J, Aubé M. 2022 LSPDD: Lamp Spectral Power Distribution Database https://lspdd.org [last accessed 28 May 2022].

[29] CIE, Commision Internationale de l'Éclairage. CIE 1988 2° SpectralLuminous efficiency function for photopic vision. Vienna: Bureau Central de la CIE; 1990.

[30] BIPM, Bureau International des Poids et Mesures (BIPM), Le Système International d'unités / The International System of Units, 9th edition, BIPM, Sèvres, France, 2019. https://www.bipm.org/utils/common/pdf/si-brochure/SI-Brochure-9.pdf [Last accessed 28 June 2022].

[31] Sánchez de Miguel A, Kyba CCM, Zamorano J, Gallego J, Gaston KJ. The nature of the diffuse light near cities detected in nighttime satellite imagery, Scientific Reports 2020;10:7829 https://doi.org/10.1038/s41598-020-64673-2

[32] Kocifaj M, Bará S. Aerosol characterization using satellite remote sensing of light pollution sources at night. Monthly Notices of the Royal Astronomical Society: Letters 2020;495:L76–L80. https://doi.org/10.1093/mnrasl/slaa060

[33] Kocifaj M, Bará S. Diffuse light around cities: new perspectives in satellite remote sensing of nighttime aerosols. Atmospheric Research 2022;266:105969 https://doi.org/10.1016/j.atmosres.2021.105969

[34] Coesfeld J, Kuester T, Kuechly HU, Kyba CCM. Reducing Variability and Removing Natural Light from Nighttime Satellite Imagery: A Case Study Using the VIIRS DNB. Sensors. 2020; 20(11):3287. https://doi.org/10.3390/s20113287

[35] Li X, Ma R, Zhang Q, Li D, Liu S, He T, Zhao L. Anisotropic characteristic of artificial light at night – Systematic investigation with VIIRS DNB multi-temporal observations. Remote Sensing of Environment 2019;233:111357. https://doi.org/10.1016/j.rse.2019.111357

[36] Wang Z, Román MO, Kalb VL, Miller SD, Zhang J, Shrestha RM. Quantifying uncertainties in nighttime light retrievals from Suomi-NPP and NOAA-20 VIIRS Day/Night Band data. Remote Sensing of Environment 2021;263:112557. https://doi.org/10.1016/j.rse.2021.112557

[37] Masana E, Carrasco JM, Bará S, Ribas SJ. A multi-band map of the natural night sky brightness including Gaia and Hipparcos integrated starlight. Monthly Notices of the Royal Astronomical Society 2021;501:5443–5456. https://doi.org/10.1093/mnras/staa4005

[38] Kasten F, Young AT. Revised optical air mass tables and approximation formula. Applied Optics 1989;28(22):4735-4738. https://doi.org/10.1364/AO.28.004735

[39] Teillet PPM. Rayleigh optical depth comparisons from various sources, Appl. Opt. 1990;29:1897-1900. https://doi.org/10.1364/AO.29.001897

[40] Earth Observation Group, 2022. Nighttime Light. Payne Institute, https://eogdata.mines.edu/products/vnl/ [Last accessed 28 June 2022].

[41] Chen H, Sun C, Chen X, Chiang K, Xiong X. On-orbit Calibration and Performance of S-NPP VIIRS DNB, in Earth Observing Missions and Sensors: Development, Implementation, and Characterization IV, edited by Xiaoxiong J. Xiong, Saji Abraham Kuriakose, Toshiyoshi Kimura, Proc. of SPIE (2016) Vol. 9881, 98812B. https://doi.org/10.1117/12.2225105

[42] Bohlin RC, Gordon KC, Tremblay PE. Techniques and Review of Absolute Flux Calibration from the Ultraviolet to the Mid-Infrared. Publications of the Astronomical Society of the Pacific, 2014;942:711-732. https://doi.org/10.1086/677655

[43] Kokaly RF, Clark RN, Swayze GA, Livo KE, Hoefen TM, Pearson NC, Wise RA, Benzel WM, Lowers HA, Driscoll RL, Klein AJ. 2017, USGS Spectral Library Version 7 Data: U.S.

Geological Survey data release, https://dx.doi.org/10.5066/F7RR1WDJ. [Last accessed 28 June 2022].

[44] IGN, 2022, Cartographic Register, Boundaries and Geographic names, Municipal boundaty lines, https://www.ign.es/web/en/ign/portal/rcc-area-rcc [last accessed, May 28, 2022]

[45] Kyba C, Stare J. 2020, Radiance Light Trends. App. Version 1.0.5 https://lighttrends.lightpollutionmap.info [last accessed, May 29th, 2022]

[46] Elvidge CD, Zhizhin M, Keith D, Miller SD, Hsu FC, Ghosh T, Anderson SJ, Monrad CK, Bazilian M, Taneja J, Sutton PC, Barentine J, Kowalik WS, Kyba CCM, Pack DW, Hammerling D. The VIIRS Day/Night Band: A Flicker Meter in Space? Remote Sens. 2022;14:1316. https://doi.org/10.3390/rs14061316